# Surface profiles with zero and finite adhesion force and adhesion instabilities

Valentin L. Popov

Technische Universität Berlin, Str. des 17. Juni 135, 10623 Berlin, Germany

**Abstract**. A simple but general analysis of the stability of axis-symmetric adhesive contacts is provided. Adhesion is considered in the JKR-approximation. Depending on the shape of the contacting bodies, various scenarios are possible, including vanishing adhesive force, complete contact as well as transitions between these extremes.

## 1. Introduction

Neutral bodies are known to attract each other via van-der-Waals forces. These forces lead to a finite work which is needed to detach two surfaces from each other, which we will call the "work of adhesion". The work of adhesion per unit area of contacting bodies is denoted as $\Delta\gamma$. The adhesive contact problem has been solved in the classical paper by Johnson, Kendall and Roberts (1971) for parabolic profiles and by Kendall (1971) for a flat cylindrical indenter.

Analysis of adhesive contacts mostly concentrates on finding the "force of adhesion" corresponding to an unstable configuration, after which no further equilibrium exists and the adhesive contact "breaks down". However, in adhesive systems another kind of instability is possible, which till now did not attract much attention: an unstable transition to the state of complete contact. Johnson (1995) has considered this instability for the case of slightly wavy surfaces. He has shown that there exist critical indentation depths at which the contact becomes unstable and the adhesive contact propagates until the surfaces come into complete contact.

Here we present a simple general discussion of both kinds of instabilities for axially symmetric contacts.

The simplest solution of the adhesive contact for arbitrary rotationally symmetric shapes with compact contact area is provided by the Method of Dimensionality Reduction [Popov und Heß (2015), (2014)]. Let us shortly recapitulate the MDR solution.

We consider a frictionless adhesive contact between two elastic bodies with Young's moduli $E_1$ and $E_2$ and Poisson numbers $\nu_1$ and $\nu_2$ and differential profile $\tilde{z} = f(r)$, where $r$ is the polar radius in the contact plane. The MDR procedure consists of the following steps:

- First, the three-dimensional profile $\tilde{z} = f(r)$ is replaced by an equivalent MDR profile

$$g(x) = |x| \int_0^{|x|} \frac{f'(r)}{\sqrt{x^2 - r^2}} dr . \qquad (1.1)$$

- The elastic bodies are replaced by an elastic foundation consisting of independent springs placed with a small spacing $\Delta x$ and having the normal stiffness

$$\Delta k_z = E^* \Delta x , \qquad (1.2)$$

where

$$\frac{1}{E^*} = \frac{1-\nu_1^2}{E_1} + \frac{1-\nu_2^2}{E_2} . \qquad (1.3)$$

- The profile $g(x)$ is pressed into the elastic foundation with the normal force $F_N$. The springs at the boundary of the contact are detached when their elongation reaches the critical value

$$\Delta l(a) = \sqrt{\frac{2\pi a \Delta\gamma}{E^*}} \qquad (1.4)$$

(rule of Heß, [Heß (2010)]).

The theorems of the MDR state that the dependencies of the three quantities $(F_N, d, a)$ (normal force, indentation depth and contact radius) in the equilibrium state reproduce exactly the solution of the original three-dimensional problem.

From the described procedure it is easy to see that the detachment criterion for the outer springs in the MDR model reads

$$d = g(a) - \Delta l(a) \tag{1.5}$$

The normal force is determined by the equation

$$F_N = 2E^* \int_0^a [d - g(x)] \, dx. \tag{1.6}$$

## 2. Discussion of Equation (1.5)

Equations (1.5) and (1.6) solve the adhesive contact problem. Equation (1.5) can be reorganized as

$$g(a) - d = \Delta l(a). \tag{1.7}$$

If the inequality

$$g(a) - d > \Delta l(a) \tag{1.8}$$

is fulfilled, then the contact radius will decrease. In the opposite case

$$g(a) - d < \Delta l(a) \tag{1.9}$$

it will increase. For the case of $d = 0$, this is illustrated in Fig. 1a.

Let us stress that in the following analysis we confine ourselves to the case of *controlled indentation*: it is assumed that the changes in non-equilibrium contact radius do not change the indentation depth. The opposite case of *controlled force* can be considered similarly.

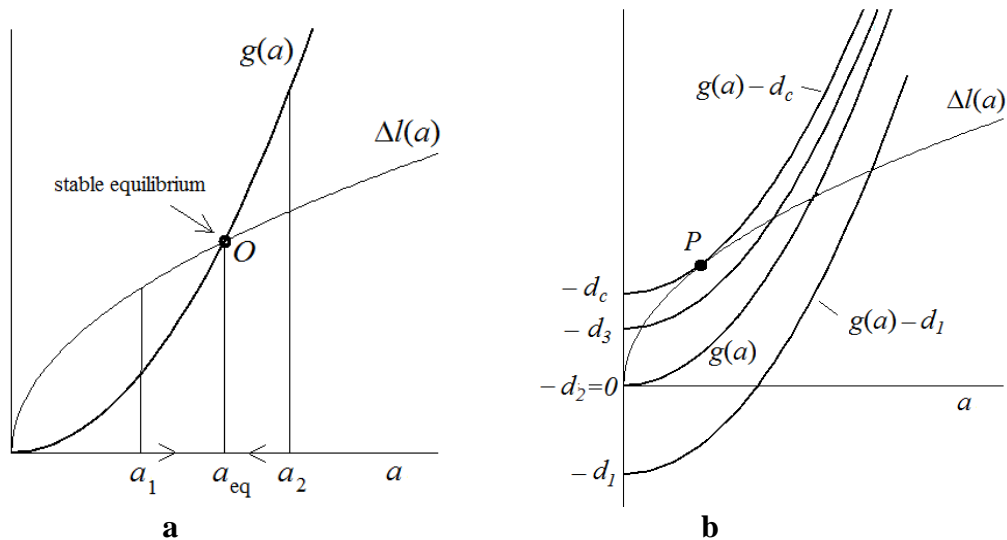

**Fig. 1** This figure illustrates the Eq. (1.7) and the inequalities (1.8) and (1.9). (a) The case $d = 0$. If the current contact radius is $a_1$, then $\Delta l > g$ and the radius will increase. If the system starts with the radius $a_2$, then $\Delta l < g$ an the radius decreases. Thus, the point *O* corresponds to a state of stable equilibrium. (b) Illustrates the case of non-vanishing – either positive or negative – indentation depth. Indentation shifts the curve of $g(a)$ either downwards (for positive indentation depths) or upwards (for negative indentation depths). The point *P* is the last one for which there exists an equilibrium state of the system.

In the general case of arbitrary indentation, the above inequalities (1.8) and (1.9) are illustrated in Fig. 1b. If the profile is indented into the medium, the cross-section point is shifted to the right and the contact radius increases. If it is pulled out of the medium, the radius of the equilibrium state shrinks. However, an equilibrium only exists for separations smaller than the critical separation $d_c$. The point $P$ can be determined from the condition

$$\frac{dg(a)}{da} = \frac{d\Delta l(a)}{da} = \sqrt{\frac{\pi \Delta \gamma}{2E^* a}}. \tag{1.10}$$

Let us discuss in a more detail the "generic case" of some intermediate separation for which the curves $g(a) - d$ and $\Delta l(a)$ have the form shown in Fig. 2.

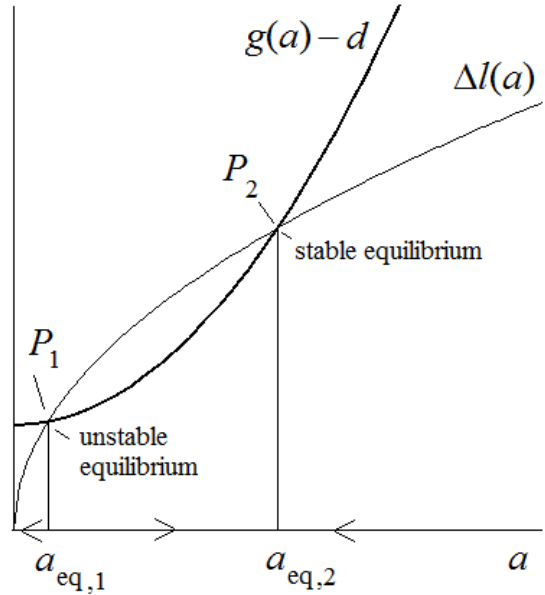

**Fig. 2** A diagram for the case of a negative but still subcritical indentation depth. Arrows indicate the direction of change of a non-equilibrium contact radius for the given indentation depth. One can see that the point $P_1$ corresponds to an unstable equilibrium and the point $P_2$ to a state of stable equilibrium.

Now one can easily see that for radii $a > a_{eq,2}$, $\Delta l(a) < g(a) - d$ and the contact will shrink. The same is valid for the region to the left of $a_{eq,1}$. Between $a_{eq,1}$ and $a_{eq,2}$, $\Delta l(a) > g(a) - d$ and the contact expands. Thus, the equilibrium point $P_1$ corresponds to an unstable equilibrium and $P_2$ to stable equilibrium.

## 3. Non-adhesion and full contact states

The above consideration is qualitatively correct for usual profile shapes which can be vaguely characterized as "convex profiles". Concave profiles may have different properties. To this category belong in particular sharp pointed profiles having the shape

$$f(r) = Cr^n, \quad \text{with } 0 < n < 1/2. \tag{1.11}$$

The diagram $\{\Delta l(a); g(a) - d\}$ has now the qualitative form shown in Fig. 3.

For $d = 0$, the curves have only one intersection point $O$ at a non-vanishing contact radius $a_O$. This only equilibrium point is *unstable*: If the initial radius is smaller than $a_O$, then the radius shrinks to zero and if it is larger than $a_O$, then it expands to infinity. This remains true for any negative indentation depth. This means that the adhesion force is in this case exactly zero. However, for *small positive* indentations, there exists a finite contact radius which is different from that of the non-adhesion problem. In this sense, the whole problem still remains "adhesive" despite the vanishing adhesive force. For large enough indentation depth, the critical state is achieved (point $P$) after which there is no further equilibrium state and the contact radius expands infinitely. The condition for this instability coincides with (1.10). As a result of the instability, the whole shape comes into complete contact.

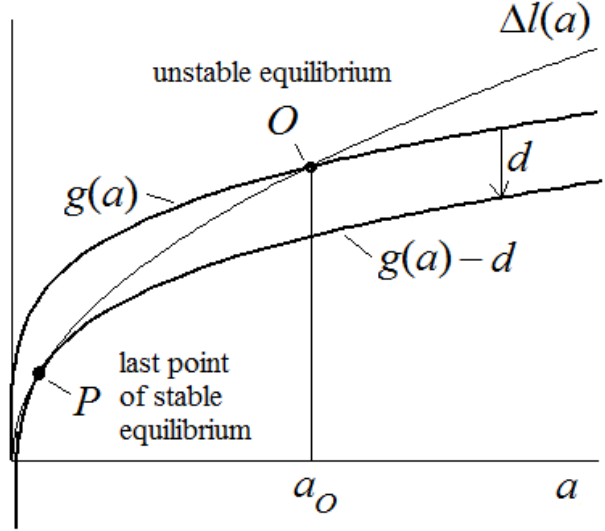

**Fig. 3** Adhesion properties of sharp pointed profiles. A power-law function $f(r) \propto r^{1/4}$ was used for illustration.

## 4. Example of jumping to complete contact for a power-law profile

We consider an axially-symmetric profile in the form of a power function given by (1.11). The MDR-transformed shape is, according to (1.1),

$$g(x) = \kappa_n f(|x|) = \kappa_n C |x|^n \quad \text{with} \quad \kappa_n = \frac{\sqrt{\pi}}{2} \frac{n\Gamma(\frac{n}{2})}{\Gamma(\frac{n+1}{2})}. \tag{1.12}$$

Equation (1.5) has the form

$$d = g(a) - \Delta \ell(a) = \kappa_n C a^n - \sqrt{\frac{2\pi a \Delta \gamma}{E^*}} \tag{1.13}$$

and the instability condition (1.10) reads

$$n\kappa_n C a^{n-1} = \sqrt{\frac{\pi \Delta \gamma}{2 E^* a}}. \tag{1.14}$$

Resolving this equation with respect to $a$ gives

$$a_c = \left( \frac{\pi \Delta \gamma}{2 E^* n^2 \kappa_n^2 C^2} \right)^{\frac{1}{1-2n}}. \tag{1.15}$$

For the normal force we get

$$F_N(a) = 2E^* \int_0^a [d - g(x)] dx = 2E^* \frac{n}{n+1} \kappa_n C a^{n+1} - \sqrt{8\pi a^3 E^* \Delta \gamma}. \tag{1.16}$$

Substitution of the critical value (1.15) for $a$ provides the critical normal force $F_{N,c}$:

$$F_{N,c} = -\frac{2(2n+1)}{n+1} (E^*)^{-\frac{n-2}{1-2n}} \left( \frac{\pi \Delta \gamma}{2} \right)^{-\frac{n+1}{1-2n}} \left( \frac{1}{n c \kappa_n} \right)^{-\frac{3}{1-2n}}. \tag{1.17}$$

This force has to be applied to the contact to produce the instability of spontaneous transition to the state of complete contact.

In the special case $n = 1/2$, we have

$$g(a) = \kappa_{1/2} C a^{1/2}, \quad \text{with } \kappa_{1/2} = \frac{1}{4} \frac{\pi^{3/2} \sqrt{2}}{\Gamma(3/4)^2} \approx 1,311,$$

$$\Delta \ell(a) = \left(\frac{2\pi \Delta \gamma}{E^*}\right)^{1/2} a^{1/2}$$
(1.18)

In the state of incipient contact, $d = 0$,

I. $\quad g(a) > \Delta l(a), \quad \text{if } C > 0,7628 \left(\frac{2\pi \Delta \gamma}{E^*}\right)^{1/2}$

II. $\quad g(a) < \Delta l(a), \quad \text{if } C < 0,7628 \left(\frac{2\pi \Delta \gamma}{E^*}\right)^{1/2}$
(1.19)

In the first case, the radius decreases until it vanishes. In the second case, it increases until complete contact is achieved.

We note once again, that the whole analysis in this paper is valid under conditions of controlled indentation depth.

## 5. Discussion

The condition for adhesive instabilities can be simply treated graphically by presenting the dependencies $g(a) - d$ and $\Delta l(a)$ in the same graph. The only prerequisite for the applicability of this procedure is the knowledge of the MDR-transformed profile $g(a)$. The condition for the instability is just the condition of touching of the curves $g(a) - d$ and $\Delta l(a)$. Depending on whether the touching is from the inner side or from outer side of the dependency $\Delta l(a)$, this leads to jump-like increase or decrease of the adhesive contact radius. We consider a number of simple cases. Of course, more complicated cases are possible, as e.g. the case of a parabolic indenter with waviness first considered by Guduru (2007), see Fig. 4.

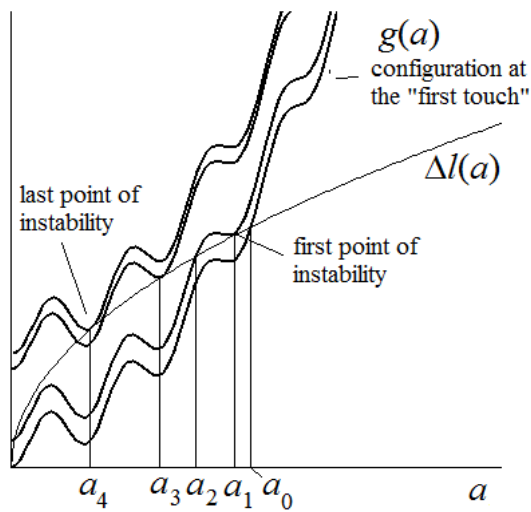

**Fig. 4 Instability analysis for a parabolic indenter with a slight waviness.**

Let us assume that initially there existed a contact with radius $a_0$ at $d = 0$. If we now pull the indenter, then the first touching of the curves will occur in the point denoted as "first point of instability". At this moment, the contact radius jumps from $a_1$ to $a_2$. The second touching occurs when the contact radius is $a_3$ and the last one at the radius $a_4$. After this, the contact radius jumps to the zero, thus the contact is broken down.